\documentclass[]{spie}  %>>> use for US letter paper
\usepackage[]{aas_macros}
\usepackage{wrapfig}
\usepackage{amsmath,amsfonts,amssymb}
\usepackage{graphicx}
\usepackage[colorlinks=true, allcolors=blue]{hyperref}

\newcommand{\rd}{\mathrm{d}}

\def\bH{{\bf H}}

\def\rk{\mathrm{k}}

\def\rm{\mathrm{m}}
\def\bm{\boldsymbol{\mathrm{m}}}

\def\rp{\mathrm{p}}

\def\br{\boldsymbol{r}}

\def\rT{\mathrm{T}}
\def\ru{\mathrm{u}}

\def\bx{{\bf x}}
\def\by{{\bf y}}

\def\balpha{\boldsymbol{\alpha}}

\def\brho{\boldsymbol{\rho}}

\title{Finding the Dark Hole with the Lights On: A New Approach to Focal Plane Wavefront Sensing}

\author[*]{Richard A. Frazin}
\affil[*]{Dept. of Climate and Space Sciences, University of Michigan, Ann Arbor, MI 48109}

\authorinfo{Email: rfrazin {\it at} umich.edu}

\begin{document} 
\maketitle

\begin{abstract}

In direct imaging of exoplanets from space, achieving the required dynamic range (i.e., planet-to-star contrast in brightness) currently relies on coronagraphic technology combined with active control of one or more deformable mirrors (DMs) to create a dark region in the image plane, sometimes called a ``dark hole.''
While many algorithms have been proposed for this purpose, all of them employ focal plane wavefront sensing (FPWS) in order to calculate the optimal DM configuration to create the desired dark hole.
All current algorithms are limited by their own success in that, as the dark hole is achieved, the FPWS procedure becomes shot-noise limited due to he low intensity in the dark hole.
This article proposes a FPWS procedure that allows determination of the optimal DM configuration without relying on information obtained when the DM is near the optimal configuration.
This article gives regression procedures for FPWS that do not assume the DM step size is small, which should allow two important improvements to the control loop: 1) performing informative FPWS observations with DM configurations that are sufficiently distant from the optimal dark hole configuration to mitigate shot-noise limitations, and 2) more accurately predicting the DM configuration that will achieve the desired objective in the dark hole control loop.
In order to treat this more challenging FPWS problem, two different representations are presented.
The first of these, is called the empirical Green's function (EGF), is easy to implement, and has a block-diagonal matrix structure that is well-suited to parallel processing.
The other representation, based on an explicit aberration expansion (EAE) requires the regression to estimate a smaller number of parameters than the EGF, but leads to a dense matrix structure.
The EGF and EAE methods both simultaneously estimate the planetary image.

\end{abstract}

\section{INTRODUCTION}\label{Intro}

Direct imaging of exoplanetary systems is difficult due to the high contrast in brightness between the planet and the star, which results in the planetary light being buried beneath details of telescope's point spread function (PSF) that can change on a large range of time-scales, ranging from minutes to days.
At this time, the scientific community is planning to perform direct imaging of exoplanets with telescopes in space, with the first such effort being NASA's WFIRST mission, which will use coronagraphic optics to suppress the starlight [{\it http://wfirst.gsfc.nasa.gov/science/presentations/vugraphs/SDT\_Jul\_2012.pdf}].
Even with a coronagraph, time-variable aberrations in the optical system (mostly due to thermal stresses) are substantial enough to necessitate the use of active optical elements, namely deformable mirrors (DMs), to create a region in the image plane that is dark enough to meet the mission requirements.
Many methods for using the DM to create the dark region, sometimes called a ``dark hole,'' have been proposed in the years following the initial idea, which is attributed to Malbet et al. in 1995.\cite{Malbet_EFC95}
In order to calculate the required DM deformation, all of these methods (reviewed in [\citenum{Kasdin_EFC16b}]) employ a focal plane wavefront sensing (FPWS) technique in which the DM implements a series ``probe shapes'' in order to determine the electric value in image plane.
One fundamental limitation in this approach is that as the DM approaches the required to make a dark hole, the hole does indeed become quite dark, so much so that the measurements needed to make the hole still darker are shot-noise limited due to the small intensity.  
One way to overcome this limitation is to determine the electric field without relying on measurements in which the DM is in a dark hole configuration.
This is not possible with current FPWS methods because they assume that they do not model the optical system sufficiently accurately to allow large DM steps and assume that the product of the phase induced by the DM step and the phase caused by unknown aberration is small.\cite{Kasdin_EFC16b}
The regression framework presented in this article models the unknown aberration in the optical system in such a way that this limitation on the step-size is removed, at least in theory (the remaining step-size limitation is DM calibration and reproducibility).

In the initial paper on the subject in 1995, Malbet et al. assumed that the unknown aberration in the optical system can be represented as equivalent aberration in a pupil plane upstream of the DM.\cite{Malbet_EFC95}
Unfortunately,  is an oversimplification for optical systems that have aberration downstream of the DM.
One optical system that is very likely to have aberrations downstream of the DM is the coronagraph for NASA's WFIRST mission, the beam interacts with well over a dozen optical surfaces after bouncing off of ``DM1,'' so there is a high probability that the equivalent aberration will vary as a consequence of a probe command sent to DM1. \cite{WFIRST_layout15}

This paper uses rigorous physical optics arguments to find the term required to represent the aberrations downstream of the DM and show that this term is of the same order as Malbet et al.'s aberration.  
After demonstrating the need for this additional term, regression equations are derived for two different representations of the post-DM aberrations, and their relative merits are presented.
Finally, this article compares this FPWS approach to those that have been published and implemented previously.

\section{Propagation Equations}

First, this section introduces some notation and formalism.  The equations in this section will be within paradigm of scalar fields. 
Vector generalizations to handle polarizing optical systems of some equations are given by [\citenum{Frazin16a}], but this important issue will be deferred to a sequel to this article.

Let $\lambda$\ be the central wavelength of the quasi-monochromatic light, and $u(\br)$\ be the analytical signal representing the electric field [\citenum{Born&Wolf,StatisticalOptics,Frazin16a}] in some plane with two-dimensional (2D) coordinate vector $\br$.
Note that, consistently with the formalism established in [\citenum{Frazin16a}], the high-frequency factor $\exp(-j 2\pi \nu t)$\ (where the speed of light $c = \lambda \nu$) has been suppressed.
The time-dependence due to statistical coherence properties of the field and will not be carried in the notation, either.

\begin{figure}[h]
\includegraphics[width=.49\linewidth,clip=]{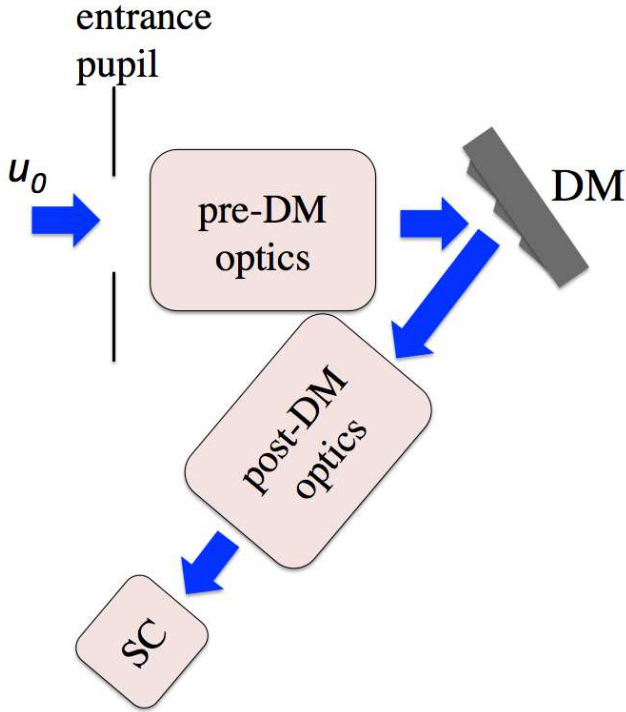}
\caption{\small Schematic diagram showing the initial field $u_0$\ impinging on the telescope entrance pupil, passing through the pre-DM optical train, reflecting off the deformable mirror (DM), passing through the post-DM optical train and, finally, forming an image in the science camera (SC).
The arrows indication the direction of light propagation.}
\label{fig_schematic}
\vspace{-4mm}
\end{figure}

Consider an optical system on a space-based platform, such as in Fig.~\ref{fig_schematic}, in which the entrance pupil is designated as plane $0$.
Assuming the angular size of the star is small enough to allow treatment as a point source, the field arising from the star hosting the putative planetary impinging on the entrance pupil is [\citenum{Frazin16a}]:
\begin{equation}
u_{\star 0}(\br_0) = \sqrt{I_\star \,} \exp(j k \balpha_\star \cdot \br  ) \,  
\end{equation}
where $k = 2 \pi / \lambda$ and $\balpha_\star$ is the 2D sky angle, in units of radians, of the (presumably small) pointing error, and $I_\star$\ is the star's irradiance.
On the other hand, the entrance pupil is presumably large enough to at least partially resolve the planetary system, so it is therefore not possible to write a general expression for the planetary field impinging on the telescope entrance pupil (unless it is assumed to be composed of unresolved point sources).
Instead, all that is available is the mutual coherence function, as provided by the well-known Van Cittert-Zernike theorem [\citenum{StatisticalOptics,Frazin16a}]:
\begin{equation}
\gamma_{\rp 0}(\br_0,\br_0') 
  =  \frac{1}{\pi}  \int_\rp   \rd \balpha \,
S(\balpha)  \exp \left[ j k \big(  \br_0 - \br_0' \big) \cdot \balpha  \right] \, ,
\label{ZvC}
\end{equation}
where $\balpha$\ is the 2D sky angle (radian units), $S(\balpha)$\ is the radiance of planetary system, and the integration is carried out over the angular extent of the planetary system, but not including the star itself (practically, this will be enforced by not trying to determine the image within some radius of the pointing center).
Eq.~(\ref{ZvC}) states that the mutual coherence of the light arriving at the Earth is proportional to the Fourier transform of the radiance emerging from the planetary system, $S(\balpha)$.  
The objective of the science observations is to estimate $S(\balpha)$, which is difficult because the stellar irradiance is much brighter than the planetary irradiance, i.e., $I_\star \gg \int_\rp \rd \balpha \, S(\balpha)$.

The optical system shown schematically in Fig.~\ref{fig_schematic} is intended to represent a space-based telescope as it does not include atmospheric modulation of light.
It is assumed to contain only a single DM, as generalization to the multiple DM case is straightforward, at least at the theoretical level presented in this article.
The light from the entrance pupil (whose plane is assigned the index number $0$) passes through the part of the optical bench called the ``pre-DM optics,'' reflects off the DM, denoted with index $d$, then passes through ``post-DM optics'' (presumably including a coronagraph), finally forming an image on the science camera (SC), whose detector surface will be given assigned index $c$.
Let the operator $\Upsilon_{d,0}\big(\br_d,\br_0 \big)$\ (where $\br_d$\ and $\br_0$\ are the 2D coordinates in the $d$ and $0$ planes, respectively) propagate the field from the telescope entrance pupil to just before the DM, thus including all effects of the pre-DM optics.
The stellar field impinging on the DM is given by
\begin{equation}
u_d(\br_d) = \Upsilon_{d,0}\big(\br_d,\br_0 \big)  u_{\star 0}(\br_0) \, ,
\label{u_d}
\end{equation}
in which integration over the $\br_0$\ coordinate is implied by the notation.
Now, the DM imposes a phase shift of $\zeta(\br_d,\bm)$\ on the beam, where the second argument $\bm$\ indicates that the DM shape ($\zeta$) is function of the DM command vector $\bm$.
Here it will be assumed that $\bm$ changes on a time-scale that is much shorter than $\tau_\rd$, the dynamical time-scale on which the spacecraft structure undergoes thermal relaxations and so on.
In this treatment, changes on the time-scale $\tau_\rd$\ are not included and quantities that vary on that time-scale are formally considered to be constant.
As a practical matter, in order to deal with changes on the time-scale $\tau_\rd$, the regressions presented below will need to updated regularly, most likely in a Kalman-filtering framework.\cite{Kasdin_EFC13}

Here, we will assume that $\zeta(\br_d,\bm)$\ is a purely real-valued, which does not allow the DM to have amplitude effects (lifting this restriction is relatively straightfoward). 
The field just after reflecting off the DM is then given by
\begin{equation}
u_d^+(\br_d,\bm) = u_d(\br_d) \exp  [j \zeta(\br_d,\bm)]
= \exp  [j \zeta(\br_d,t)]  \Upsilon_{d,0}\big(\br_d,\br_0 \big)  u_{\star 0}(\br_0) \, ,
\label{u_d+}
\end{equation}
where the $^+$\ superscript emphasizes that the light has just reflected off of DM.
Similarly, $\Upsilon_{c,d}\big( \br_c,\br_d\big)$,  where $\br_c$\ is the 2D coordinate in the SC plane, propagates $u_d^+$\ from the DM to the SC.
The propagator  $\Upsilon_{c,d}$\ includes the post-DM optical train.
Note, in a break with the conventions established in [\citenum{Frazin16a}],  $\Upsilon_{c,d}$\ does not include interaction with the surface specified by index $d$, which, in this case, is taken into account by the factor $\exp  [j \zeta(\br_d,\bm)]$.
The, the field in the SC arising from the starlight is given by
\begin{align}
u_c(\br_c,\bm) & = \Upsilon_{c,d}\big( \br_c,\br_d\big) u_d^+(\br_d) \nonumber \\ 
& =  \Upsilon_{c,d}\big( \br_c,\br_d\big) \exp  [j \zeta(\br_d,\bm)]  \Upsilon_{d,0}\big(\br_d,\br_0 \big)  u_{\star 0}(\br_0) \, .
\label{u_c}
\end{align}
Under this formalism, Eq.~(\ref{u_c}) implies integration not only over the pupil plane coordinate $\br_0$, but also over the intermediate coordinate in the DM plane, $\br_d$.

Due to unknown and time-variable (on the $\tau_\rd$\ time-scale) aberrations throughout the entire optical system, the propagators  $ \Upsilon_{c,d}$  and $\Upsilon_{d,0}$ are only partially known.
It is helpful to decompose $ \Upsilon_{c,d}$  and $\Upsilon_{d,0}$\ into known (denoted with superscript $^\rk$) and unknown (superscript $^\ru$) as follows:
\begin{align}
 \Upsilon_{c,d}\big( \br_c,\br_d\big) & =  \Upsilon_{c,d}^\rk\big( \br_c,\br_d\big) +  \Upsilon_{c,d}^\ru\big( \br_c,\br_d\big)  
\label{cd-ku}  \\
\Upsilon_{d,0}\big(\br_d,\br_0 \big) & = \Upsilon_{d,0}^\rk\big(\br_d,\br_0 \big) + \Upsilon_{d,0}^\ru\big(\br_d,\br_0 \big) \, ,
\label{d0-ku}
\end{align}
in which the known propagators $\Upsilon_{d,0}^\rk$ and $ \Upsilon_{c,d}^\rk$\ must be implemented with numerical integrations.
The unknown propagators  $ \Upsilon_{c,d}^\ru$  and $\Upsilon_{d,0}^\ru$ account for the unknown aberrations in the post-DM and pre-DM optical trains, respectively.
Using Eqs.~(\ref{cd-ku}) and (\ref{d0-ku}) in Eq.~(\ref{u_c}) results in an expression for the field consisting of known and unknown parts (using the same superscripting convention), i.e., 
\begin{equation}
u_c(\br_c,\bm) = u_c^\rk(\br_c,\bm) + u_c^\ru(\br_c,\bm) \, ,
\label{u_c-ku}
\end{equation}
where
\begin{equation}
 u_c^\rk(\br_c,\bm) = 
   \Upsilon_{c,d}^\rk\big( \br_c,\br_d\big) \exp  [j \zeta(\br_d,\bm)]  \Upsilon_{d,0}^\rk\big(\br_d,\br_0 \big) 
 u_{\star 0}(\br_0)  \, , 
\label{u_c^k}
\end{equation}
and
\begin{multline}
 u_c^\ru(\br_c,\bm) = \big\{
 \Upsilon_{c,d}^\rk\big( \br_c,\br_d\big) \exp  [j \zeta(\br_d,\bm)]  \Upsilon_{d,0}^\ru\big(\br_d,\br_0 \big) \; + \\
 \Upsilon_{c,d}^\ru\big( \br_c,\br_d\big) \exp  [j \zeta(\br_d,\bm)]  \Upsilon_{d,0}^\rk\big(\br_d,\br_0 \big) \;  + \; 
 \Upsilon_{c,d}^\ru\big( \br_c,\br_d\big) \exp  [j \zeta(\br_d,\bm)]  \Upsilon_{d,0}^\ru\big(\br_d,\br_0 \big) 
\big\}  u_{\star 0}(\br_0)
\label{u_c^u-1}
\end{multline}
It is helpful to define the unknown quantity $g^\ru(\br_d)$, called the {\it effective pre-DM aberration}:
\begin{equation}
g^\ru(\br_d) \equiv  \Upsilon_{d,0}^\ru\big(\br_d,\br_0 \big)  u_{\star 0}(\br_0) \, .
\label{g}
\end{equation}
Eq.~(\ref{g}) states that all of the aberrations in the pre-DM optics result in an unknown function of the coordinate in the DM plane, $g^\ru(\br_d)$, and is essentially the same as the aberration originally assumed by Malbet et al.,\cite{Malbet_EFC95} as will be explained below in Sec.~\ref{Previous}.
Note that $g^\ru(\br_d) $\ also includes any unknown aberration in the DM itself that varies on the time-scale $\tau_\rd$, which, unfortunately excludes uncalibrated deformations of the DM shape that vary as its configuration changes.
Using Eq.~(\ref{g}) to simplify Eq.~(\ref{u_c^u-1}), one obtains:
\begin{multline}
 u_c^\ru(\br_c,\bm) = 
\Upsilon_{c,d}^\rk\big( \br_c,\br_d\big) \exp  [j \zeta(\br_d,\bm)] g^\ru(\br_d) \: + \\
\Upsilon_{c,d}^\ru\big( \br_c,\br_d\big) \exp  [j \zeta(\br_d,\bm)]  \Upsilon_{d,0}^\rk\big(\br_d,\br_0 \big) u_{\star 0}(\br_0) +
\Upsilon_{c,d}^\ru\big( \br_c,\br_d\big) \exp  [j \zeta(\br_d,\bm)] g^\ru(\br_d) \, .
\label{u_c^u}
\end{multline}
One can see that $g^\ru$ and $\Upsilon_{c,d}^\ru$\ should be of roughly equal importance, as the aberrated field $u_c$\ in Eq.~(\ref{u_c^u}) has terms that linear in each one, as well as a final term that bi-linear in the two quantities.   
Below, in Sec.~\ref{Sec_EAE}, it will be seen that the importance of the  $\Upsilon_{c,d}^\ru$\ term should be proportional to the number optical surfaces in the post-DM optical train (or at least the number of fully illuminated surfaces that are upstream of the coronagraph).

\section{Science Camera Intensity}

The intensity of the light impinging on the SC is given by the sum of the planetary intensity and the stellar intensity:
\begin{equation}
I_c(\brho, \bm) = I_{\rp c}(\brho,\bm) \: + \: I_c(\brho,\bm) \, 
\end{equation}
where $\brho \equiv \br_c$\ is the coordinate in the SC plane.
Now, the planetary intensity $I_{\rp c}$\ is vastly fainter than the stellar intensity and a relatively simple model of the optical system, with no unknown aberration, should be perfectly adequate for calculating the planetary contribution.
Let the needed simplified propagation operator be denoted by $\Upsilon_\rp(\brho,\br_0,\bm)$.   
Choosing $\Upsilon_\rp(\brho,\br_0,\bm) =  \Upsilon_{c,d}^\rk\big( \brho,\br_d\big) \exp  [j \zeta(\br_d,\bm)]  \Upsilon_{d,0}^\rk\big(\br_d,\br_0 \big) $\ should certainly be a valid option, but $\Upsilon_{c,d}^\rk$ and $ \Upsilon_{d,0}^\rk $\ operators may be unnecessarily computationally expensive, so that cheaper approximations may serve for calculating the planetary contribution.
Any approximation should include the effect of the DM ($ \exp  [j \zeta(\br_d,\bm)] $), as it will modulate the planetary light.
In any event, the planetary intensity on the SC is given by integrating over the mutual coherence function in Eq.~(\ref{ZvC}) [\citenum{StatisticalOptics,Frazin16a}]:
\begin{align}
I_{\rp c}(\brho,\bm) &= \Upsilon_\rp(\brho,\br_0,\bm) \Upsilon_\rp^*(\brho,\br_0',\bm) \gamma_{\rp 0}(\br_0,\br_0') \nonumber \\
& =  \Upsilon_\rp(\brho,\br_0,\bm) \Upsilon_\rp^*(\brho,\br_0',\bm)  \frac{1}{\pi}  \int_\rp   \rd \balpha \,
S(\balpha)  \exp \left[ j k \big(  \br_0 - \br_0' \big) \cdot \balpha  \right]  \nonumber \\
& =  \frac{1}{\pi}  \int_\rp   \rd \balpha \, S(\balpha) \, \bigg\{
 \Upsilon_\rp(\brho,\br_0,\bm) \Upsilon_\rp^*(\brho,\br_0',\bm)  \exp \left[ j k \big(  \br_0 - \br_0' \big) \cdot \balpha  \right] 
\bigg\} \, ,
\label{planet-intensity}
\end{align}
where the reader is reminded that that Eq.~(\ref{planet-intensity}) is, in fact, a triple integral as the propagation operators imply integration over $\br_0$ and $\br_0'$.
The quantity inside the braces is called the {\it planetary intensity kernel} [\citenum{Frazin13,Frazin16a}].
In the absence of the confounding starlight, estimating $S(\balpha)$\ amounts to solving a multi-frame deconvolution problem.
In this deconvolution problem, the coronagraphic optics make the point-spread function (PSF) spatially variant, at least close to the inner working angle, and the PSF depends on the DM command vector $\bm$, as well.

The stellar contribution to the intensity is given by $ I_c(\brho,\bm) = u_c(\brho,\bm) u_c^*(\brho,\bm)$, using Eq.~(\ref{u_c-ku}) it can be written as:
\begin{equation}
 I_c(\brho,\bm) -  I_0(\brho,\bm)  =  u_c^\rk(\brho,\bm)  u_c^{\ru*}(\brho,\bm) + u_c^\ru(\brho,\bm)  u_c^{\rk*}(\brho,\bm) +  u_c^\ru(\brho,\bm)  u_c^{\ru *}(\brho,\bm) \, ,
\label{I_c-ku}
\end{equation}
where the known portion of the intensity, $ I_0(\brho,\bm) \equiv u_c^\rk(\brho,\bm)  u_c^{\rk *}(\brho,\bm)$, has been moved to left-hand-side of the equation.
The small, unknown aberrations in the optical system are taken into account by the unknown function $g^\ru$\ and the unknown operator $\Upsilon_{c,d}^\ru$.
Without defining precise metrics to be more quantitative, $g^\ru$, defined in Eq.~(\ref{g}), is presumably smaller than $\Upsilon_{d,0}^\rk\big(\br_d,\br_0 \big) u_{\star 0}(\br_0)$\ and $\Upsilon_{c,d}^\ru$\ should be smaller than $\Upsilon_{c,d}^\rk$, so that $u_c^\ru$\ in Eq.~(\ref{u_c^u}) contains terms that are of 1\underline{st} and 2\underline{nd} order in these unknown quantities.   
It then follows that the right-hand-side of Eq.~(\ref{I_c-ku}) contains terms that of 1\underline{st}, 2\underline{nd}, 3\underline{rd} and 4\underline{th} order in these small, unknown quantities.
Below, only the 1\underline{st} order terms are given, as deriving the others is straightforward, and they can be treated by standard procedures involving linearization and iteration, if necessary.
Then,
\begin{align}
& \: \: \: \: \:  \: \: \: \: \: I_c(\brho,\bm) - I_0(\brho,\bm) \: \: \: \approx \nonumber \\
&u_c^\rk(\brho,\bm) \big\{
\Upsilon_{c,d}^{\rk*}\big( \br_c,\br_d\big) \exp  [-j \zeta(\br_d,\bm)] g^{\ru*}(\br_d)  +
\Upsilon_{c,d}^{\ru*}\big( \br_c,\br_d\big) \exp  [-j \zeta(\br_d,\bm)]  \Upsilon_{d,0}^{\rk*}\big(\br_d,\br_0 \big) u_{\star 0}^*(\br_0) 
\big\}  \nonumber \\
+ & u_c^{\rk *}(\brho,\bm) \big\{
\Upsilon_{c,d}^\rk\big( \br_c,\br_d\big) \exp  [j \zeta(\br_d,\bm)] g^\ru(\br_d)  + 
\Upsilon_{c,d}^\ru\big( \br_c,\br_d\big) \exp  [j \zeta(\br_d,\bm)]  \Upsilon_{d,0}^\rk\big(\br_d,\br_0 \big) u_{\star 0}(\br_0) 
\big\} \, ,
\label{I_c-1st}
\end{align}
where the reader will recall that the phase imparted by the DM, $\zeta(\br_d,\bm)$, was taken to be real.

The science camera measurement is subject to the effects of noise, both due to photon counting statistics [\citenum{StatisticalOptics}] (shot noise) and readout noise, and perhaps other effects.   
Fortunately, there is a new generation of ultra-low noise IR cameras capable of kHz readouts, such as the SWIR single photon detector, SAPHIRA  eAPD and the MKIDS \cite{SWIR_detector14,SELEX_APD12,Saphira_eAPD14,Mazin_MKIDS14}, making the issue of readout noise far less critical.
Taking the noise into account, the measured value of the science camera intensity is obtained by summing the stellar and planetary contributions:
\begin{equation}
I_\rm(\brho_l,\bm_i) = (\Delta t_i) I_c(\brho_l,\bm_i) +(\Delta t_i)  I_{\rp c}(\brho_l,\bm_i) + \nu(\brho_l,t_i) \, ,
\label{I_m}
\end{equation}
where $\brho_l$\ is the postion of the $l$\underline{th} SC detector pixel, $\bm_i$\ is the $i$\underline{th} DM command vector, $t_i$\ is the time-stamp associated with $\bm_i$, $\Delta t_i$\ is the corresponding exposure time, and $\nu(\brho_l,t_i)$\ is a sample of the random process that describes the detector noise.  
Eq.~(\ref{I_m}) makes several assumptions:
\begin{itemize}
\item{The detector pixels are small enough so that integration of $I_c(\brho,\bm)$ and $ I_{\rp c}(\brho,\bm)$\ over the pixel area is not necessary.  Implementing such an integration is straightforward.}
\item{The noise, $ \nu(\brho_l,t_i)$\ has no important dependence on the unknown quantities $S(\balpha)$,  $g^\ru(\br_d)$\ and the unknown operator $\Upsilon_{c,d}^\ru(\brho_l,\br_d)$. 
This cannot be strictly true as the shot noise will depend on these values, especially as the success is achieved in creating a dark hole.
In an iterative procedure, the statistics of $\nu$\ can be calculated with current estimates of unknowns, if necessary. 
However, the entire point of the FPWS strategy presented here is to gain the required information from DM configurations in which shot noise is not a big problem in the first place.}

\end{itemize}

\section{Regression Strategies}

The regression procedure estimates the effective pre-DM aberration $g^\ru$, the post-DM aberrations $\Upsilon_{c,d}^\ru$ and the planetary image $S$.
Most of the information pertaining to $g^\ru$\ and $\Upsilon_{c,d}^\ru$\ comes from instances when the DM is not making a dark hole, while the estimate of $S$\ depends most heavily on the instances when the hole is darkest. 
The objective of this section is to give the regression formulae for estimating these quantities.

Since $S(\balpha)$\ and $g^\ru(\br_d)$\ are continuous functions on some portion of $\mathbb{R}^2$ and $\mathbb{C}^2$, respectively, they must be parameterized somehow for computation.
The most natural way to achieve this is via series expansions, which reduces the unknown continuous quantities to a finite set of expansion coefficients.   
The image processing community has a vast literature on how best to do this, including multi-resolution representations (e.g., wavelets), but these issues will not be treated here, and instead, only generic forms will be given.
Estimating the kernel of the propagation operator  $\Upsilon_{c,d}^\ru(\brho_l,\br_d)$\ is more challenging because it defines a continuous mapping  $\Upsilon_{c,d}^\ru : \, \mathbb{C}^2 \rightarrow \mathbb{R}^2$.
Below, two strategies for estimating $\Upsilon_{c,d}^\ru$\ will be presented.
The first approach is called the {\it Empirical Green's Function}, and despite its high dimensionality, it is (relatively) straightforward to implement and very well-suited to parallel computation.
The EGF was first introduced by the author in [\citenum{Frazin16_spie}] in the context of exoplanet imaging from ground-based observatories.
The second approach, called {\it Explicit Aberration Expansion}, requires challenging and expensive Fresnel propagation computations and is much less easy to parallelize. 
However, in most situations, it will likely require the estimation of many fewer unknown quantities than the EGF and therefore should make more efficient use of observational resources, at least in theory.

The regression can expressed in linear algebra terms via the canonical equation $\by = \bH \bx$, where $\by$\ is derived from the observations (SC pixel values), $\bx$\ is a vector of unknown regression coefficients to be estimated, and $\bH$\ is the model-based matrix that relates the two.
Generalization to a Kalman filtering framework, as in [\citenum{Kasdin_EFC13}], is left to future work.
The vector $\bx$\ is composed of three sub-vectors as follows:
\begin{equation}
\bx = 
\left[\begin{array}{l}
\bx_\Upsilon \\ \bx_g \\ \bx_\rp
\end{array}\right]  \, ,
\label{x}
\end{equation}
where $\bx_\Upsilon$\ contains the regression coefficients associated with estimating $\Upsilon_{c,d}^\ru$, and $\bx_g$ and $\bx_\rp$\ contain the coefficients specifying the effective pre-DM aberration $g^\ru$\ and the planetary image $S$, respectively.
Correspondingly, the matrix $\bH$\ is partitioned into three submatrices as:
\begin{equation}
\bH = 
\big[
\bH_\Upsilon \: \: \bH_g  \: \: \bH_\rp
\big]  \, ,
\label{x}
\end{equation}
where the subscripts have meanings that correspond those in Eq.~(\ref{x}).

The vector $\by$\ is modeled by Eq.~(\ref{I_m}), and is arranged so that each subvector $\by_l$\ corresponds to the time-series associated with the $l$\underline{th} SC pixel.
Recalling that the time-stamp $t_i$\ corresponds to the DM command $\bm_i$
$by$\ is given by:
\begin{equation}
\by = 
\left[\begin{array}{l}
\by_0 \\ \vdots \\ \by_l \\ \vdots \\ \by_{N-1}
\end{array}\right]  \, ,
\label{y}
\end{equation}
where $N$\ is the number of SC pixels and 
\begin{equation}
 \by_l = 
\left[
\begin{array}{c} 
I_\rm(\brho_l,\bm_0) - I_0(\brho_l,\bm_0)   - \nu(\brho_l,t_0)  \\
\vdots \\
I_\rm(\brho_l,\bm_i) - I_0(\brho_l,\bm_i)  - \nu(\brho_l,t_i)    \\
\vdots \\
I_\rm(\brho_l,\bm_{T-1}) - I_0(\brho_l,\bm_{T-1}) - \nu(\brho_l,t_{T-1})    
\end{array}  
 \right] \, ,
 \label{y_l}
\end{equation}
in which $T$\ is the number of DM steps used in the estimation.
Thus, $\by_l$\ has $T$\ elements and $\by$\ has $NT$\ elements.
The unknown effective pre-DM aberration $g^\ru(\br)$ can be approximated with the following expansion:
\begin{equation}
g^\ru(\br) \approx \sum_{l=0}^{N_g-1}  \big( a_l + jb_l  \big) \psi_l^g(\br) \, ,
\label{g-expansion}
\end{equation}
where $N_g$\ is the number of terms, and the $\{ \psi_l^g(\br) \}$\ are real-valued expansion functions (e.g., annular Zernike polynomials).
Since $g^\ru(\br)$ is complex-valued, the expansion has a real part given by the expansion coefficients $\{ a_l \}$, and the imaginary part is given by the expansion coefficients $\{ b_l \}$. 
The vector $\bx_g$\ is given by:
\begin{equation}
\bx_g^\rT = [a_0, \hdots , a_l , \hdots , a_{N_g-1}, b_0, \hdots , b_l , \hdots , b_{N_g-1}]^\rT \, ,
\label{x_g}  
\end{equation}
where the superscript $^\rT$\ indicates tranposition (in this case, to make it into a column vector).
The corresponding elements of the $NT \times 2N_g$\ matrix $\bH_g$\ are calculated by inserting Eq.~(\ref{g-expansion}) into Eq.~(\ref{I_c-1st}) and performing the requisite integrations for each DM position $\bm_i$\ and SC pixel position $\brho_l$.
Similarly, the planetary image $S(\balpha)$\ can be approximated with a series expansion as follows:
\begin{equation}
S(\balpha) \approx \sum_{l=0}^{N_\rp-1}  p_l \psi_l^\rp(\balpha) \, ,
\label{planet-expansion}
\end{equation}
where $N_\rp$\ is the number of terms included in the expansion, the $\{ p_l \}$\ are the real-valued expansion coefficients, and the $\{ \psi_l^\rp (\balpha) \}$\ are the real-valued expansion functions used for describing the planetary image.
The vector $\bx_\rp$\ in Eq.~(\ref{x}) holds the planetary coefficients and is given by:
 \begin{equation}
\bx_\rp^\rT = [p_0, \hdots , p_l , \hdots , p_{N_\rp-1}]^\rT \, .
\label{x_p}  
\end{equation}
The corresponding elements of the $NT \times N_\rp$\ matrix $\bH_\rp$\ are calculated by inserting Eq.~(\ref{planet-expansion}) into Eq.~(\ref{planet-intensity}) and performing the needed integrations.

\subsection{Empirical Green's Function}\label{Sec_EGF}

The operator $\Upsilon_{c,d}(\brho,\br)$\ is Green's function that solves diffraction problem with the field propagating from sources in plane $d$\ to plane $c$.\cite{Born&Wolf}
$\Upsilon_{c,d}^\rk(\brho,\br)$\ is the known part of this operator, and the part that must be determined from the measurements, $\Upsilon_{c,d}^\ru(\brho,\br)$, will be called the Empirical Green's Function (EGF).
When treating polarization effects, it has a straightforward generalization to vector fields, called the {\it Empirical Green's Tensor (EGT)}, which will be explained in the followup to this article [it was introduced in \citenum{Frazin16_spie}].

At any fixed position in the detector plane, $\brho_l$\ the kernel of the operator $\Upsilon_{c,d}^\ru(\brho_l,\br) $\ is function on some region of $\mathbb{R}^2$ (as is  $g^\ru(\br)$).   
Then, at any given $\brho_l$, it also admits a series expansion:
\begin{equation}
\Upsilon_{c,d}^\ru(\brho_l,\br) \approx \sum_{k=0}^{M_l-1}  \big( c_{l,k} + jd_{l,k}  \big) \psi_k^l(\br) \, ,
\label{EGF-expansion}
\end{equation}
where $M_l$\ is the number of terms in the series corresponding to the $l$\underline{th} detector pixel, the $\{\psi_k^l \}$\ are the expansion functions corresponding to the $l$\underline{th} detector pixel.
Eq.~(\ref{EGF-expansion}) is called the EGF expansion.
Since $\Upsilon_{c,d}^\ru(\brho_l,\br)$ is complex-valued, the expansion has a real part given by the expansion coefficients $\{ c_{l,k} \}$\ and the imaginary part is given by the expansion coefficients $\{ d_{l,k} \}$.
It is important to emphasize that, from this point of view, $\Upsilon_{c,d}^\ru(\brho_l,\br)$ and $\Upsilon_{c,d}^\ru(\brho_m,\br)$\ are completely independent functions of $\br$\ unless $l=m$, so that the choices made for the $\{\psi_k^l \}$\ (as well as $M_l$) can be made independently for each SC pixel positon.
It is, in fact, this independence that makes implementation of the EGF so easy to parallelize, as one obtains an independent regression for each pixel position $l$.
This independence corresponds to a block-diagonal structure in linear algebra terms, as will be demonstrated directly.
The vector $\bx_\Upsilon$\ associated with the EGF is composed of $N$\ (one for each SC pixel) sub-vectors as follows:
\begin{equation}
\bx_\Upsilon^\rT = [\bx_0, \hdots , \bx_l , \hdots , \bx_{N-1}]^\rT \, ,
\label{x_Ups}  
\end{equation}
in which
\begin{equation}
\bx_l = [c_{l,0}, \hdots , c_{l,k} , \hdots , c_{l,M_l-1}, d_{l,0}, \hdots , d_{l,k} , \hdots , d_{l,M_l-1}]^\rT \, ,
\label{x_l}  
\end{equation}
which as $2M_l$\ components.
By creating an entirely separate expansion series for each SC pixel, Eq.~(\ref{EGF-expansion}) assures that the matrix $\bH_\Upsilon$\ will have a block-diagonal representation.
Let $\bH_{\Upsilon,l}$\ be the $T \times 2M_l$\ block corresponding to the sub-vector $\bx_l$\ in Eq.~(\ref{x_l}).
Its elements are calculated by inserting Eq.~(\ref{EGF-expansion}) into Eq.~(\ref{I_c-1st}).
The resulting linear system has the form:
\begin{equation}
\left[\begin{array}{l}
\by_0 \\ \vdots \\ \by_l \\ \vdots \\ \by_{N-1}
\end{array}\right] 
=
\left[\begin{array}{l l l l l l l}
\bH_{\Upsilon,0} &             &           &              &                & \bH_{g,0}     & \bH_{\rp,0} \\
           & \ddots &           &              &                & \vdots       & \vdots \\
           &             & \bH_{\Upsilon,l} &              &                & \bH_{g,l}     & \bH_{\rp,l} \\
           &             &           & \ddots  &                & \vdots       & \vdots \\
           &             &           &              & \bH_{\Upsilon,N-1} & \bH_{g,N-1} &\bH_{\rp,N-1} 
\end{array}\right] 
\left[\begin{array}{l}
\bx_\Upsilon \\ \bx_g \\ \bx_\rp
\end{array}\right] \, ,
\label{EGF-block}
\end{equation}
in which $\bH_{g,l}$\ is the portion of the $\bH_g$\ matrix associated with the $l$\underline{th} SC pixel, and similarly for $\bH_{\rp,l}$.

The partial block-diagonal structure in Eq.~(\ref{EGF-block}) immediately suggests an iterative, alternating minimization procedure in which $\bx_g$\ and $\bx_\rp$\ are held constant while estimate of $\bx_\Upsilon$ is updated, and vice-versa (perhaps using algorithms that enjoy improved convergence rates by employing ``costs-to-move'').   
Indeed, the planetary coefficients $\bx_\rp$\ can probably be assumed to 0, except in pixels that happen to be inside a dark hole.

The price to be payed for the ``parallelizability'' and relative simplicity of the EGF approach is the fact specifying the EGF requires estimating $2M_l$\ free parameters for each SC pixel.
It may well be practical to solve for a sufficient number of coefficients to allow a rather high-order expansion in Eq.~(\ref{EGF-expansion}), especially if high-cadence detectors are available.
As an example, assume the expansion functions $\{ \psi_{k}^l(\br) \}$\ are given by the annular Zernike polynomials up to 20\underline{th} order, which would correspond to a total of $230$\ polynomials, or $M_l = 230$.   
Since we must solve for the real and imaginary parts of the expansion coefficients, the vector $\bx_l$ would have $460$ components that must be estimated from the regression.   
As a rather arbitrary example, assume that the cadence of the SC is 0.01 seconds, and that each exposure corresponds to a different DM command $\bm$.
After 100 seconds, each SC pixel would have $10^4$\ observations from which to estimate these 460 coefficients, overdetermining the problem by a factor of about 20.   
The number of coefficients that one may determine is likely to be limited by the inversion of a matrix of size $2 M_l \times 2 M_l $ (computation time proportional to $M_l^3$), or $460 \times 460$\ in this example.   Note that the author's desktop machine was able to invert a $1000 \times 1000$ matrix of random numbers (which tends to result in a poorly conditioned system) in about 0.1 s.

\subsection{Explicit Aberration Expansion}\label{Sec_EAE}

The explicit aberration expansion (EAE), in which the propagator $\Upsilon_{c,d}^\ru(\brho,\br)$\ is assumed to be the result of unknown aberration functions in $P$\ planes.  
The EAE was introduced in [\citenum{Frazin16a}] in the context of ground-based imaging.
The fact that the EAE should require estimating fewer parameters than the EGF can be seen as follows:
In the EAE, each aberration function requires $M_l=M$\ expansion coefficients to be estimated, then there would be a total of $PM$\ in the $\bx_\Upsilon$ vector, whereas, in the EGF, the $\bx_\Upsilon$ would contain $NM$\ elements.
One would expect $PM \ll NM$ since the number of planes $P$ is much smaller than the number of SC pixels $N$.
Furthermore, in some optical systems the needed value of $P$ can be reduced by treating aberrations in conjugate planes an equivalent aberration in one plane, as will be shown below.  
The disadvantages of the EAE relative to the EGF are:
\begin{itemize}
\item{the first order approximation, which prohibits aberrated fields interacting with other aberrations (much like the Born approximation in scattering theory), whereas the EGF assumes only that the optical system is linear in the field, not the aberrations}
\item{the difficult Fresnel integrations needed to propagate the field from one plane to the next}
\item{the fact that the matrix $\bH_\Upsilon$\ is dense, instead of block-diagonal, as it is in the EGF.}
\end{itemize}

To derive the EAE, assume that the post-DM optics contain unknown aberrations in $P$\ planes, with the first plane in the post-DM optics containing aberration given the index $1$\ (the index $0$\ was already used to signify the telescope entrance pupil), and the final aberrated plane having the index $P$.   
Let the propagator between plane $l$\ and plane $l+1$\ be denoted by $\Upsilon_{l+1,l}(\br_{l+1},\br_l)$, which includes interaction with the $l$\ surface but not the $l+1$ surface.
The only exception to this rule are propagators that start at the DM plane $d$ (note that the is already included in $g^\ru$), so  $\Upsilon_{1,d}(\br_{1},\br_d) = \Upsilon_{1,d}^k(\br_{1},\br_d)$\ is a known operator.

Following the earlier procedure, the propagator $\Upsilon_{l+1,l}$\ can be decomposed into known and unknown parts:
\begin{align}
\Upsilon_{l+1,l}(\br_{l+1},\br_l) & =  \Upsilon_{l+1,l}^\rk(\br_{l+1},\br_l) + \Upsilon_{l+1,l}^\ru(\br_{l+1},\br_l) \nonumber \\
& = \Upsilon_{l+1,l}^\rk(\br_{l+1},\br_l)\big[ 1 + A_l^\ru(\br_l)  \big]   \, ,
\label{Ups_l-total}
\end{align}
which assumes that he unknown character of the propagator is confined to an unknown aberration function in plane $l$, given by the complex-valued function $A_l^\ru(\br_l)$.
While it may be tempting to set $ A_l^\ru(\br_l) = \exp[ j \phi_l(\br_l)]$\ for some (possibly complex-valued) phase aberration function $\phi_l(\br_l)$\ as was done in [\citenum{Frazin16a}], it is of little practical value as one must immediately perform the Taylor expansion of the exponential $ A_l^\ru(\br_l) \approx 1 + j  \phi_l(\br_l) -  \phi_l^2(\br_l)/2 + \cdots$\ and deal the with consequences of the approximation.
On the other hand, this problem is avoided by treating $A_l^\ru$\ as the sum of two functions with real and imaginary parts, i.e., $A_l^\ru(\br_l) = A_l^{\ru r}(\br_l) + j A_l^{\ru i}(\br_l) $.
The case of a small, real-valued phase perturbation $\phi_l(\br_l)$\ is recovered by fixing the value  $A_l^{\ru r}(\br_l) = 1$\ and setting $ A_l^{\ru i}(\br_l) = \phi_l(\br_l) $.

The known field leaving the DM, i.e., excluding the pre-DM aberrations is:
\begin{equation} u_d^\rk(\br_d,\bm) = \exp  [j \zeta(\br_d,\bm)]  \Upsilon_{d,0}^\rk\big(\br_d,\br_0 \big)  u_{\star 0}(\br_0) \, ,
\label{u_d^k}
\end{equation}
 and using Eq.~(\ref{Ups_l-total}), the contribution to the aberrated field, $u_C^\ru$, arising from the aberration in plane $l$ arriving at SC is: 
\begin{equation}
 \Upsilon_{c,l+1}^\rk(\br_c,\br_{l+1}) \Upsilon_{l+1,l}^\rk(\br_{l+1},\br_l)  A_l^\ru(\br_l)  \Upsilon_{l,d}^\rk(\br_l,\br_d)   u_d^\rk(\br_d,\bm) \, ,
\label{u-ab-l+1}
\end{equation} 
where $\Upsilon_{l,d}^\rk$\ is a known operator that propagates the field from the $d$ plane to the $l$ plane that is given by contraction:
\begin{equation}
\Upsilon_{l,d}^\rk(\br_l,\br_d) = \left[\prod_{k=1}^{l-1} \Upsilon_{k+1,d}^\rk(\br_{k+1},\br_l) \right]  \Upsilon_{1,d}^\rk(\br_{1},\br_d) \, .
\label{Ups_ld}
\end{equation}
Similarly, $\Upsilon_{c,l}^\rk(\br_c,\br_l) = \Upsilon_{c,l+1}^\rk(\br_c,\br_{l+1}) \Upsilon_{l+1,l}^\rk(\br_{l+1},\br_l) $,
so that total contribution of the post-DM aberrations to the SC field is, to 1\underline{st} order in the aberrations,
\begin{equation}
 \sum_{l=1}^{P} \Upsilon_{c,l}^\rk(\br_c,\br_l)  A_l^\ru(\br_l)  \Upsilon_{l,d}^\rk(\br_l,\br_d)   u_d^\rk(\br_d,\bm) \, .
\label{u_c^u-abex-1}
\end{equation}
Thus, the unknown propagator $\Upsilon_{c,d}^\ru$ is given by:
\begin{equation}
\Upsilon_{c,d}^\ru \big(\br_c,\br_d   \big) \approx \sum_{l=1}^{P} \Upsilon_{c,l}^\rk(\br_c,\br_l)  A_l^\ru(\br_l)  \Upsilon_{l,d}^\rk(\br_l,\br_d)   \, ,
\label{EAE}
\end{equation}
where the strict equality is lost due to the fact that terms in which aberrated fields do not interact with downstream aberrations are excluded.
This should be permissible if $|A_l^\ru(\br_l)| \ll 1$, as per Eq.~(\ref{Ups_l-total}).
Eq.~(\ref{EAE}) is the EAE.
One obvious consequence is that the value $\Upsilon_{c,d}^\ru$\ is roughly proportional to $P$, the number of planes containing aberration, which is likely every optical surface in the post-DM optical train, or at least the number of fully illuminated planes upstream of the coronagraph.
Unlike the EGF, which has completely independent expansion for each SC position $\br_c$, in the EAE, each aberration $A_l$\ potentially influences the  value of the field at each SC position.
Note that implementation of the propagators  $ \Upsilon_{c,l}^\rk(\br_c,\br_l) $\ and  $\Upsilon_{l,d}^\rk(\br_l,\br_d) $\ in Eq.~(\ref{EAE}) will likely require Fresnel integrations, which can be expensive and challenging.

The function $A_l^\ru(\br_l)$\ can be reduced to a finite parameter set by again employing a series expansion:
\begin{equation}
A_l^\ru(\br_l) \approx \sum_{k=0}^{M_l-1}  \big( c_{l,k} + jd_{l,k}  \big) \psi_k^l(\br_l) \, ,
\label{A_l-expansion}
\end{equation}
where $M_l$\ is the number of terms in the sum, the coefficients $\{c_{l,k} \}$ and $\{d_{l,k} \}$\ are real and the $\{  \psi_k^l(\br_l) \}$\ are expansion functions.
Eq.~(\ref{A_l-expansion}) looks rather similar to Eq.~(\ref{EGF-expansion}), but, in this case, the expansion approximates the aberration function in the $l$\underline{th} plane.
As mentioned above, if one is only concerned with small, real-valued phase perturbations then one can set $\{c_{l,k} = 1 \}$.
Formally, the vector $\bx_\Upsilon$, looks like that for the EGF, but EAE will have many fewer components, as explained above.
In the EAE formulation, each subvector $\bx_l$\ corresponds to the coefficients associated with aberration $A_l^\ru$.
So,
\begin{equation}
\bx_\Upsilon^\rT = [\bx_0, \hdots , \bx_l , \hdots , \bx_{N-1}]^\rT \, , \: \mathrm{and}
\label{x_Ups-A_l}  
\end{equation}
\begin{equation}
\bx_l = [c_{l,0}, \hdots , c_{l,k} , \hdots , c_{l,M_l-1}, d_{l,0}, \hdots , d_{l,k} , \hdots , d_{l,M_l-1}]^\rT \, ,
\label{x_l-A_l}  
\end{equation}
To calculate the elements of the $\bH_\Upsilon$\ matrix associated with EAE, Eq.~(\ref{A_l-expansion}) is used in Eq.~(\ref{EAE}), which, in turn, is substituted in to Eq.~(\ref{I_c-1st}).

The resulting linear system has the form:
\begin{equation}
\left[\begin{array}{l}
\by_0 \\ \vdots \\ \by_l \\ \vdots \\ \by_{N-1}
\end{array}\right] 
=
\left[\begin{array}{l l l }
\bH_{\Upsilon,0}    & \bH_{g,0}     & \bH_{\rp,0} \\
\vdots               & \vdots       & \vdots \\
\bH_{\Upsilon,l} & \bH_{g,l}     & \bH_{\rp,l} \\
   \vdots  & \vdots       & \vdots \\
\bH_{\Upsilon,N-1} & \bH_{g,N-1} &\bH_{\rp,N-1} 
\end{array}\right] 
\left[\begin{array}{l}
\bx_\Upsilon \\ \bx_g \\ \bx_\rp
\end{array}\right] \, ,
\label{EAE-dense}
\end{equation}
which, unlike Eq.(\ref{EGF-block}), is a dense linear system that has no obvious structure to exploit.

\subsubsection{The Role of Equivalent Aberrations in the EAE}

As mentioned above, it may be permissible to reduce the number of planes included in the EAE in Eq.~(\ref{EAE}), which reduces the size of $\bx_\Upsilon$.
In short, under the approximations of geometrical optics, if aberrations are present in one or more planes that are conjugate to plane $l$, then these conjugate planes may be omitted from the EAE in Eq.~(\ref{EAE}), effectively replacing the aberrations in these planes with an {\it equivalent aberration} in plane $l$.
Demonstrating this requires the concept of equivalent aberrations, which was explained in Sec. 4.1 of [\citenum{Frazin16a}], but it is included here due to its direct relevance.

\begin{figure}[h]
\includegraphics[width=.49\linewidth,clip=]{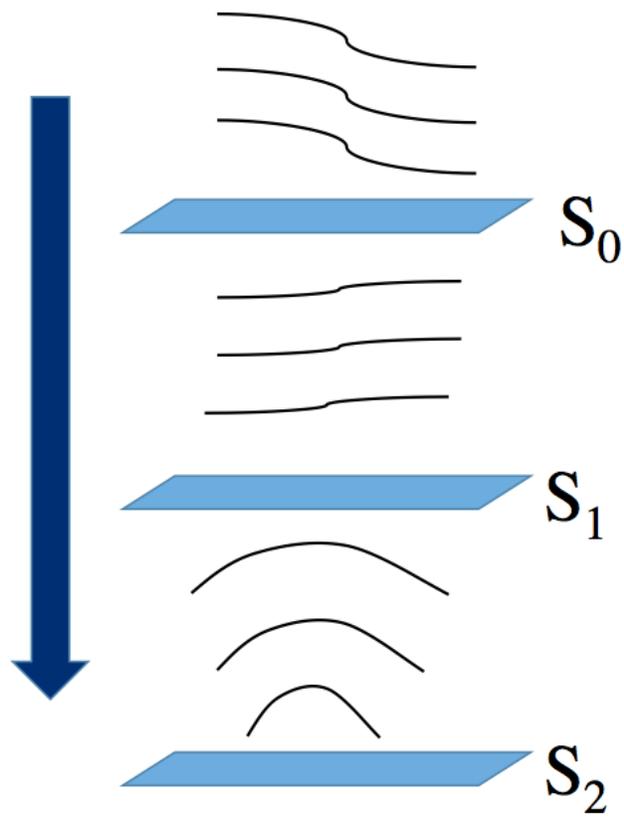}
\caption{\small Conceptual diagram depicting light passing through phase screens on surfaces $S_0$\ and $S_1$ before arriving at surface $S_2$.
The arrow indicates the direction of propagation, and the black curves indicate a sequence of wavefronts.}
\label{fig_3surfaces}
\vspace{-4mm}
\end{figure}

Consider the conceptual diagram in Fig.~\ref{fig_3surfaces}, which represents three transmitting surfaces, downstream of some DM given the command vector $\bm$, that interact with optical radiation that is propagating from top to the bottom in the picture.  The light first interacts with surface $S_0$, then propagates to the suface $S_1$, interacts with it, and finally propagates to $S_2$.
Let coordinate vectors in the surfaces $S_0$, $S_1$ and $S_2$ be given by $\br_0$, $\br_1$ and $\br_2$, respectively.
The field just before interacting with $S_0$\ is denoted by $u_0(\br_0,\bm)$, and similarly for $u_1(\br_1,\bm)$ and $u_2(\br_2,\bm)$.  
The relationship between $u_0(\br_0,\bm)$\ and $u_1(\br_1,\bm)$\ is defined by the propagation operator $\Upsilon_{1,0}\big(\br_1,\br_0\big)$, with a similar meaning for the operator $\Upsilon_{2,1}\big(\br_2,\br_1\big)$, so that 
\begin{align}
u_1(\br_1,\bm) & = \Upsilon_{1,0}\big(\br_1,\br_0\big)  u_0(\br_0,\bm) \, , \: \:  \mathrm{and} \label{u1-u0} \\
u_2(\br_2,\bm) & = \Upsilon_{2,1}\big(\br_2,\br_1\big)  u_1(\br_1,\bm) \, . \label{u2-u1} 
\end{align}

Consider a thought experiment in which the propagators $\Upsilon_{1,0}$\ and $\Upsilon_{2,1}$\ include the phase aberrations imparted by the surfaces $S_0$\ and $S_1$, denoted by $\phi_0(\br_0)$\ and $\phi_1(\br_1)$, respectively.  
Then two propagators can be written as 
\begin{align}
\Upsilon_{1,0}\big(\br_1,\br_0\big) = &\Upsilon_{1,0}^\rk\big(\br_1,\br_0\big) \exp[j \phi_0(\br_0)]  \: \: \mathrm{and} \label{u1-u0_ab} \\
\Upsilon_{2,1}\big(\br_2,\br_1\big) = & \Upsilon_{2,1}^\rk\big(\br_2,\br_1\big) \exp[j \phi_1(\br_1)] \, , \label{u2-u1_ab}
\end{align}
where $\Upsilon_{1,0}^\rk$\ and $\Upsilon_{2,1}^\rk$\ are known propagators.
Then, using   Eqs.~(\ref{u1-u0_ab}) and (\ref{u2-u1_ab}) inside Eqs.~(\ref{u1-u0}) and (\ref{u2-u1}), 
\begin{equation}
u_2(\br_2,\bm) = \Upsilon_{2,1}^\rk\big(\br_2,\br_1\big) \exp[j \phi_1(\br_1)] \Upsilon_{1,0}^\rk\big(\br_1,\br_0\big) \exp[j \phi_0(\br_0)] u_0(\br_0,\bm) \, .
\end{equation}
Now, suppose one wishes to treat the cumulative effect of the aberrations on $S_0$\ and $S_1$\ with some equivalent aberration only on $S_1$, and let this equivalent aberration be represented by $\phi_1'$.
In other words, does there exist some $\phi_1'$\ that satisfies the condition
\begin{align}
 u_2(\br_2,\bm) & =  \nonumber \\
 &\Upsilon_{2,1}^\rk\big(\br_2,\br_1\big) \exp[j \phi_1(\br_1)] \Upsilon_{1,0}^\rk\big(\br_1,\br_0\big) \exp[j \phi_0(\br_0)] u_0(\br_0,\bm)  = \nonumber \\
 &\Upsilon_{2,1}^\rk\big(\br_2,\br_1\big) \exp[j \phi_1'(\br_1,\bm)] \Upsilon_{1,0}^\rk\big(\br_1,\br_0\big)  u_0(\br_0,\bm)
\: \mathrm{?} \label{equiv-ab-question}
\end{align}
Clearly, the validity of Eq.~(\ref{equiv-ab-question}) is independent of the leftmost operator $ \Upsilon_{2,1}^\rk$, and after dropping it, $\phi_1'(\br_1,\bm)$\ must be given by:
\begin{equation}
\exp[j\phi_1'(\br_1,\bm)] = \frac{ \exp[j \phi_1(\br_1)] \Upsilon_{1,0}^\rk\big(\br_1,\br_0\big) \exp[j \phi_0(\br_0)] u_0(\br_0,\bm)}
{ \Upsilon_{1,0}^\rk\big(\br_1,\br_0\big)  u_0(\br_0,\bm) } \, .
\label{equiv-ab}
\end{equation}
Ignoring the possibility of zeros in the denominator, from Eq.~(\ref{equiv-ab}) one can see that there does indeed exist an equivalent aberration $\phi_1'$ on $S_1$\ that replaces both of the original aberrations on $S_0$ and $S_1$, however, {\it the price of this substitution is that the equivalent aberration is depends on the DM configuration ($\bm$), even though the original aberrations do not.}   Thus, the value of $\phi_1'(\br_1,\bm)$ will fluctuate as the DM modulates $u_0(\br_0,\bm)$.
However, there is (at least) one non-trivial choice for the operator $\Upsilon_{1,0}^k$\ that removes the problem of dependence on the DM configuration.   
Suppose that surfaces $S_1$\ and $S_0$\ are in conjugate planes under geometrical optics approximations.  
In that case, the propagator $\Upsilon_{1,0}$\ is given by $ \Upsilon_{1,0}^\rk\big(\br_1,\br_0\big) = \delta( \br_0 - \beta \br_1)$, where $\delta(\cdot)$\ is the Dirac delta, and $\beta$\ is some scalar magnification factor, and Eq.~(\ref{equiv-ab}) becomes (integrating over $\br_0$):
\begin{equation}
\exp[j\phi_1'(\br_1)] =  \exp j [ \phi_1(\br_1) + \phi_0(\beta \br_1)] \, ,
\label{equiv-ab-conj}
\end{equation}
where the second argument of $\phi_1'$\ ($\bm$) has been dropped since the $\bm$-dependent factors cancel, as it depends only the original aberrations $\phi_0$\ and $\phi_1$.
{\it Thus, when two planes are conjugate under geometrical optics, their aberrations can be replaced by an equivalent aberration that is independent of the incident field}.
On the other hand, Eq.~(\ref{equiv-ab-conj}) says that, under geometrical optics approximations,  aberrations in certain optical systems with many surfaces possibly can be represented by equivalent aberrations in only several planes, if the aberrations in many planes are all conjugate to only a few planes.
Solving for the equivalent aberrations under this approximation is a form of tomography and in the future it may be called ``aberration tomography.''
One should also be careful before assuming two planes are conjugate in a given optical system, especially as certain optics components such as DMs and coronagraphs can easily break conjugacy relationships.

\section{Relationship to Previous FPWS Formulations}\label{Previous}

The developments presented here allow one to understand the initial assumption in the original paper in 1995 by Malbet et al. [\citenum{Malbet_EFC95}] that has since been repeated in all subsequent work [e.g., \citenum{Traub_Nulling06, Martinache_SpeckleCancel12, Kasdin_EFC16b}].   
Under the assumption that $\Upsilon_{c,d}^\ru = 0$, i.e., that there is no unknown aberration downstream of the DM, the 2\underline{nd} and 3\underline{rd} terms of Eq.~(\ref{u_c^u}) are zero.
Then, using Eqs.~(\ref{u_c-ku}), (\ref{u_c^k}) and (\ref{u_c^u}) one obtains:
\begin{equation}
u_c(\br_c,\bm) \: \: \mathrm{``} = \mathrm{"} \: \:
 \Upsilon_{c,d}^\rk\big( \br_c,\br_d\big) \exp  [j \zeta(\br_d,\bm)] \big\{  \Upsilon_{d,0}^\rk\big(\br_d,\br_0 \big) 
 u_{\star 0}(\br_0) \; + \: g^\ru(\br_d) \big\} \, ,
\label{wrong}
\end{equation} 
which is the equivalent of, e.g., Eq.~1 of  [\citenum{Malbet_EFC95}], Eq.~2 of [\citenum{Traub_Nulling06}] or Eq.~1 of [\citenum{Kasdin_EFC16b}].
In Eq.~(\ref{wrong}), the equals sign is shown in quotation marks to emphasize the invalidity of the $\Upsilon_{c,d}^\ru = 0$ assumption.

Interestingly, even if an aberration in the post-DM optics is in a plane that is conjugate to the DM-plane or some plane in the pre-DM optics, it cannot be taken into account with the effective pre-DM aberration, $g^\ru(\br_d)$\ in Eq.~(\ref{g}).
Thus, even in this favorable situation, Eq.~(\ref{wrong}) is not valid.
To demonstrate this fact, as a thought experiment, consider the following form for $\Upsilon_{c,d}^\ru$:
\begin{equation}
\Upsilon_{c,d}^\ru\big( \br_c,\br_d\big) = \Upsilon_{c,q}^\rk\big( \br_c,\br_q\big) 
\exp [j \vartheta_q(\br_q)] \delta \big(\br_d - \br_q \big) \, ,
\label{q-ab}
\end{equation}
where $\vartheta_q(\br_q)$\ is the sole unknown aberration downstream of the DM in some plane denoted by index $q$, and $\Upsilon_{c,q}^\rk$\ is a known propagator.
The optical conjugacy between the $d$\ and $q$\ planes is expressed by using the Dirac delta function as the propagation operator between these two planes.
Then, using Eq.~(\ref{q-ab}), the 2\underline{nd} term in Eq.~(\ref{u_c^u}) becomes (the 3\underline{rd} term is product of two presumably small unknowns):
\begin{equation}
\Upsilon_{c,q}^\rk\big( \br_c,\br_q\big) 
\exp j [ \vartheta_q(\br_q) + \zeta(\br_q,\bm)]  \Upsilon_{d,0}^\rk\big(\br_q,\br_0 \big) u_{\star 0}(\br_0) \, ,
\end{equation}
which cannot be included in the $g^\ru(\br_d)$\ function defined in Eq.~(\ref{g}) since the leftmost operator, $\Upsilon_{c,q}^\rk$, integrates over the $q$-plane, not the $d$-plane as does Eq.~(\ref{wrong}).   

The reader may wonder how various groups have successfully created dark regions when starting with the incorrect Eq.~(\ref{wrong}).
As explained, e.g., in [\citenum{Kasdin_EFC16b}], these EFC methods do not attempt to estimate the unknown aberration (essentially $g^\ru$ in those formulations), rather they directly estimate the field in the image plane using DM ``probes,'' under the assumption that the probe deformations are small enough that the effect of the unknown aberration on the changes in the field induced by the DM steps are negligible (which is certainly true when steps are small enough).   
Thus, as the DM shape converges to the optimal configuration, the field in the image plane is constantly reassessed. 
One of major difficulties with this paradigm is that as the DM converges to the optimal configuration, the dark hole becomes very dark indeed, making photon counting noise (shot noise) a limiting factor in the accuracy of the technique. 
{\it In contrast, the methods presented here make no stipulation that the DM step size is small, which allows the regression to be informed by DM shapes that are relatively far from a dark hole configuration. 
Then, after the regression procedure has estimated the aberrations, the DM can be commanded to take its optimal configuration.}
Another difficulty with current methodology is that after determining the field in the image plane, the DM is commanded to take the configuration that algorithm, be it stroke-minimization, electric field conjugation, or something else, calculates to be optimal for creating the dark hole.
However, there is no guarantee that this DM step will be small enough so that the effect of unknown aberration will be negligible.
{\it In contrast, the method presented would allow these algorithms to include the effect of the aberrations whatever the step-size, making solution more accurate.  This could improve convergence to the dark hole and possibly make it even darker.}

\section{Conclusions}

The FPWS methods presented here are potentially exciting because they mitigate two important difficulties with current methodologies, namely the shot-noise limitations in the dark hole (because the regression can be informed by DM configurations that are relatively far from dark solution), and the accuracy of the optimum that is found by the minimization algorithm (because the cost functions will be accurate with larger step-sizes).  
These methods also have the capability of being used iteratively to help mitigate the effects of the linearization in the regression.   
For example, once the aberrations have been estimated, they can be included in the known propagators, so that the new unknown propagators, determined by another regression, treat only smaller aberrations.   

This article has completely neglected polarization effects, which are likely to be important, since partial linear and  circular polarization is caused by reflection off of mirrors.
This is critical because orthogonal polarizations do not interfere, so that the diffraction pattern from beam with spatially variable polarization will be different from that derived in the scalar approximation, in which of the light interferes.
The effects of polarization in telescopes have been described by [\citenum{Breckinridge15}], who shows that a telescope is better described by a point-spread-matrix (which operates on on the Stokes vector) than a scalar PSF.
The regression methods described here can be made fully polarimetric using methods similar to those in [\citenum{Frazin16a}] and in [\citenum{Frazin16_spie}], which presented the vector generalization of the empirical Green's function called the empirical Green's tensor.
The author plans to present the polarimetric versions of these equations in a followup article.
Another limitation of the proposed methods is the accuracy and reproducibility of the DM shape $\zeta(\bm)$, which leads into DM calibration issues that are beyond the scope of this discussion.
Whatever the achievable accuracy of the DM, in the end, the true value of $\zeta(\bm)$\ is a stochastic quantity, and it would be necessary to characterize some of its statistical properties in order to assess the likely consequences of DM figure error.
Demonstrating the suitability of the methods presented here for a space mission, first in simulation and then in the laboratory, is left to future efforts.

\section*{Acknowledgments} 

The author thanks Jim Breckinridge for providing comments and engaging in helpful discussions.
This work has been supported by NSF Award \#1600138 to the University of Michigan. 

%\bibliography{../JOSAA16/exop} % bibliography data in report.bib
\bibliographystyle{spiebib} % makes bibtex use spiebib.bst

\end{document}